\documentclass[letter,
               keeplastbox,   
               ]{jacow}
\usepackage{pdfpages,multirow,ragged2e} %
\makeatletter%
	\ifboolexpr{bool{xetex}}
	 {\renewcommand{\Gin@extensions}{.pdf,%
	                    .png,.jpg,.bmp,.pict,.tif,.psd,.mac,.sga,.tga,.gif,%
	                    .eps,.ps,%
	                    }}{}
\makeatother

\ifboolexpr{bool{xetex} or bool{luatex}} 
 {}                                      
 {\usepackage[utf8]{inputenc}}           

\usepackage[USenglish]{babel}

\ifboolexpr{bool{jacowbiblatex}}%
 {%
  \addbibresource{jacow-test.bib}
  \addbibresource{biblatex-examples.bib}
 }{}
\usepackage{physics, hyperref}

\begin{document}

\title{realizing steady-state microbunching with optical stochastic crystallization}

\author{M. Wallbank\thanks{wallbank@fnal.gov}, J. Jarvis\thanks{jjarvis@fnal.gov}, Fermi National Accelerator Laboratory, Batavia, IL, USA}
	
\maketitle

\begin{abstract}
  Optical Stochastic Cooling (OSC) is a state-of-the-art beam cooling technology first demonstrated in 2021 at the IOTA storage ring at Fermilab's FAST facility. A second phase of the research program is planned to run in early 2025 and will incorporate an optical amplifier to enable significantly increased cooling rates and greater operational flexibility.

  In addition to beam cooling, an OSC system can be configured to enable advanced control over the phase space of the beam. An example operational mode could enable crystallization, where the particles in a bunch are locked into a self-reinforcing, regular microstructure at the OSC fundamental wavelength; we refer to this as Optical Stochastic Crystallization (OSX).  OSX represents a new path toward Steady-State Microbunching (SSMB), which may enable light sources combining the high brightness of a free-electron laser with the high repetition rate of a storage ring.  Such a source has applications from the terahertz to the extreme ultraviolet (EUV), including high-power EUV generation for semiconductor lithography.

  The status of the OSC experimental program at IOTA and its potential to achieve the first demonstration of SSMB during the upcoming experimental run will be discussed.
\end{abstract}

\section{INTRODUCTION}

The Integrable Optics Test Accelerator (IOTA) is a 40\,m storage ring at the Fermilab Accelerator Science and Technology (FAST) facility~\cite{FAST}, which can store either 150\,MeV electrons or 2.5\,MeV protons~\cite{IOTA}.  A significant part of the research program has been focused on Optical Stochastic Cooling (OSC)~\cite{OSC-CDR}, with a world-first demonstration being achieved with 100\,MeV electrons in 2021~\cite{Jarvis2022}.

OSC is an advanced beam cooling technology which extends the widely-used stochastic cooling technique~\cite{VanDerMeer1985} from microwave to optical frequencies and bandwidths.  Since the mechanism relies on bunch sampling, by increasing the system bandwidth the cooling times can be correspondingly significantly reduced~\cite{OSC1993}.  The `transit-time' method of OSC~\cite{Zolotorev1994} implemented at IOTA employs two undulators as the `pickup'  and `kicker' elements, designed respectively to sample the longitudinal profile of the beam bunch through its emitted radiation, and to apply corrective energy kicks to the particles.  In between, the particles pass through a dispersive chicane to translate the energy deviations to a longitudinal coordinate spread, and the light is focused and optionally amplified.  The particles and their pickup radiation then co-propagate through the kicker undulator and exchange energy based on their relative phase.  Due to the periodic nature of the radiation field, the resulting energy kick for the particles is also periodic and is given by~\cite{Jarvis2022}
\begin{equation}
  \delta p / p = \kappa u(s) \sin(k_0 s),
\end{equation}
where $\kappa$ is the maximum kick value; $k_0 = 2\pi / \lambda_r$ is the radiation wavenumber for an optical wavelength $\lambda_r$; $s$ is the longitudinal displacement from the pickup to the kicker relative to the reference particle, which obtains zero kick; and $u(s)$ is an envelope function, with $u(0) = 1$ and $u(s) = 0$ for $|s| > N_u \lambda_r$ (where $N_u$ is the number of undulator periods), which accounts for the bandwidth of the integrated system.  Typically, the system is configured so particles of the design energy arrive in phase with their radiation at the kicker undulator such that they exchange no net energy; off-momentum particles either gain energy from or lose energy to the radiation fields to bring their energies toward the design value.  After many passes through the system, the beam is cooled.  The cooling mechanism relies on two important components: sufficient randomization of the particles between subsequent passes to ensure collective effects are not introduced; and sufficiently fine bunch sampling such that on average the particles feel their own radiation more than that from other particles within the bandwidth of the system at the pickup.

Steady-State Microbunching (SSMB) has been proposed as a potential future light source technology, combining the high-brightness of a free-electron laser with the high average power of a ring-based source~\cite{Ratner2010}.  The concept involves creating and sustaining microstructure within a circulating beam for the production of high-power coherent radiation.  Conventional SSMB approaches utilize a laser cavity with very high (>1\,MW) stored power to modulate the bunch at the optical wavelength.  The underlying SSMB concept was recently shown experimentally~\cite{Deng2021}, and further work is ongoing for a full demonstration~\cite{Kruschinski2024,Li2023}.

\section{OPTICAL STOCHASTIC CRYSTALLIZATION}

\begin{figure}
  \centering
  \includegraphics[width=0.95\linewidth]{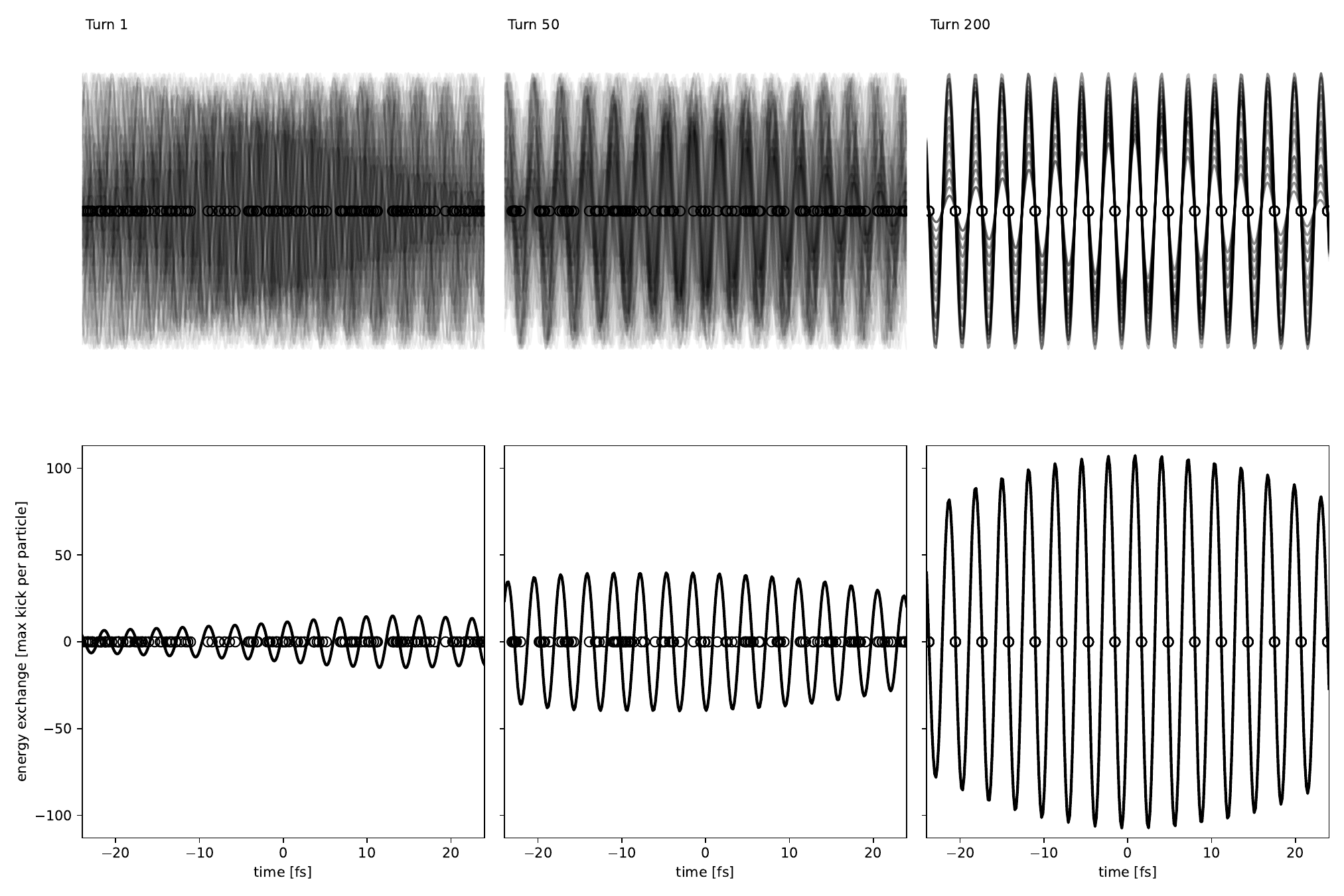}
  \caption{Demonstration of the OSX mechanism using a simple 1D simulation, with three snapshots shown left to right.  Individual particles are represented as circles with their corresponding energy kicks represented as sinusoids.  The top panels show the wakes from individual particles, while the bottom panels show the corresponding superposition of all particle wakes.  A very high system gain was applied here for demonstration purposes.}
  \label{fig:OSXMechanism}
\end{figure}

An OSC system can be utilized to enable SSMB within an appropriately configured storage ring; we refer to this form of SSMB as Optical Stochastic Crystallization (OSX).  Within an OSX setup, microstructure is formed at the fundamental optical wavelength and reinforced through the energy exchanges produced by the OSC system.  An OSX system intentionally violates the randomization requirement that is typical of stochastic cooling systems, which permits and enhances the development of microbunches through collective effects enabled by the finite system bandwidth.  As microbunches form they are strongly self-cooled through the superposition of the particle radiation fields.  A basic demonstration of the mechanism is shown in Fig.~\ref{fig:OSXMechanism}.

The requirements for enabling OSX are: sufficient optical gain in the OSC system to initiate and preserve the bunching; a sufficiently low ring momentum compaction to support the longitudinal bunch structure; matched signs of the slip factor in the OSC chicane and the ring; appropriate minimization of linear and nonlinear transverse-to-longitudinal coupling throughout the system, e.g. through minimization of the dispersion invariant in the undulators.

\begin{figure}
  \centering
  \includegraphics[width=0.95\linewidth]{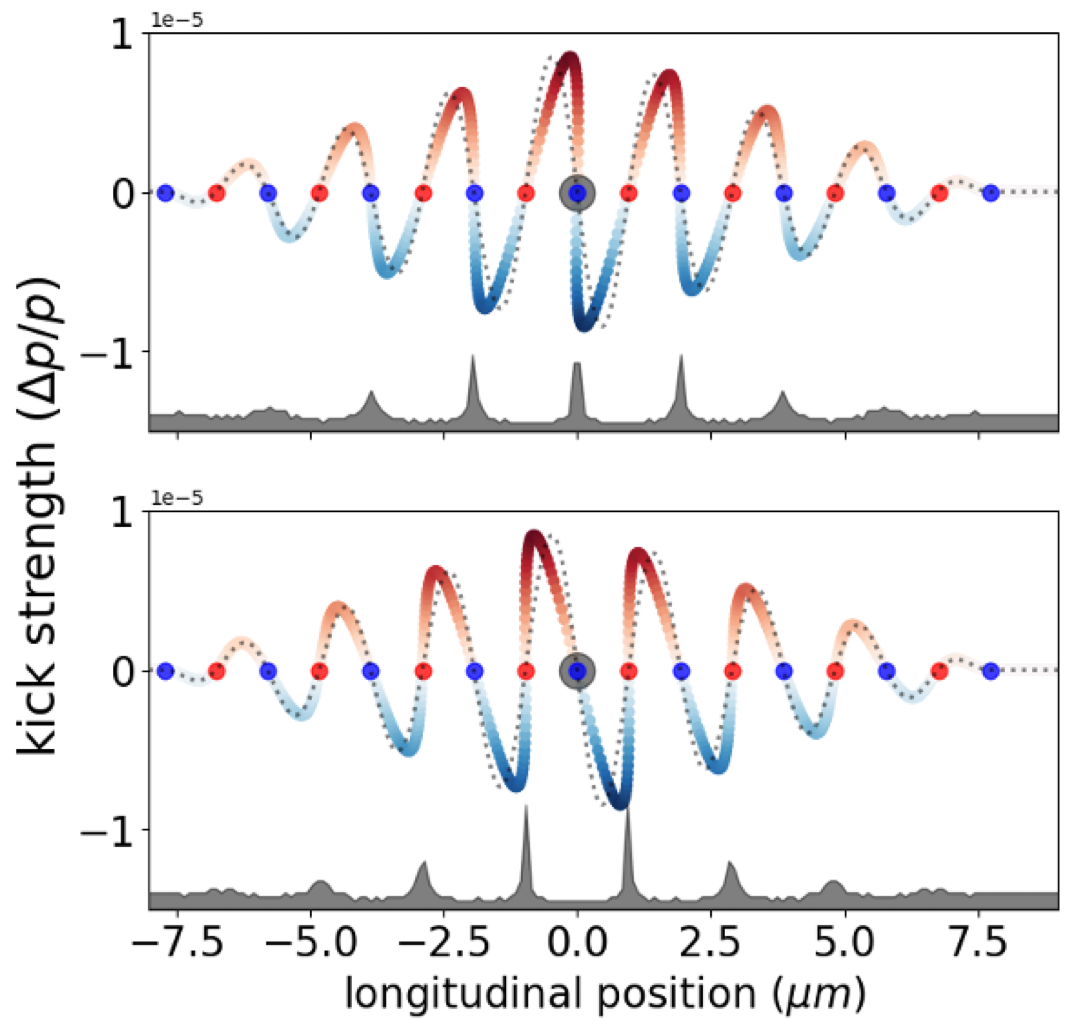}
  \caption{Impact of the sign of the ring momentum compaction for an OSC system utilizing a negative slip factor in the chicane: negative compaction (top); positive compaction (bottom).  A positive density perturbation is located at the origin and creates a sinusoidal energy modulation on the surrounding particles (dotted black line) upon transiting an OSC system.  Subsequent passage through the storage ring will reinforce (top) or reduce (bottom) the original perturbation depending on the sign of the ring momentum compaction; in this case, SSMB is sustained with a negative ring compaction.  A very high system gain was applied here for demonstration purposes.}
  \label{fig:StorageRing}
\end{figure}

Since the OSC kick is periodic, as shown in Fig.~\ref{fig:StorageRing}, a positive density perturbation at the pickup will produce a corresponding periodic energy modulation in the nearby particles at the kicker.  The OSX concept then uses the momentum compaction of the storage ring to drive the modulated particles towards cooling phases of the OSC kick from the original density perturbation.  This produces mutually reinforcing density perturbations at the OSC fundamental wavelength (Fig.~\ref{fig:StorageRing} top).  Conversely, if the sign of the compaction does not match that of the OSC bypass (Fig.~\ref{fig:StorageRing} bottom), the resulting density perturbations are driven instead towards heating phases and the initial perturbation will not be reinforced.

Longitudinal bunch lengthening from transverse coupling~\cite{Shoji2004} requires mitigation at the undulators to ensure the structure can form and be reinforced.  Preliminary simulations suggest reducing the dispersion invariant relative to the transverse emittance at this part of the ring is a viable option, and can be achieved through lattice design and beam cooling, for example using the OSC system.  Depending on the particular application, minimization of other effects may be important, such as longitudinal quantum excitation due to partial slip factor~\cite{Shoji1996,Deng2020}.

On subsequent turns through an OSX system, the energy kicks become strongly enhanced, due to the self and collective cooling, as the population and definition of the microbunches increases.  This constitutes an effective increase in the gain of the OSC system, potentially by many orders of magnitude.  Depending on the details of implementation, this may result in ultra-cold microbunches and greatly reduce the stringent requirement for ultra-low compaction that is typical of some conventional SSMB configurations.  An additional advantage of OSX-based SSMB is that the microstructure in the beam and the radiation fields are self-synchronizing and thus do not require precise phase-locking to an external laser.

\begin{figure}
  \centering
  \includegraphics[width=0.95\linewidth]{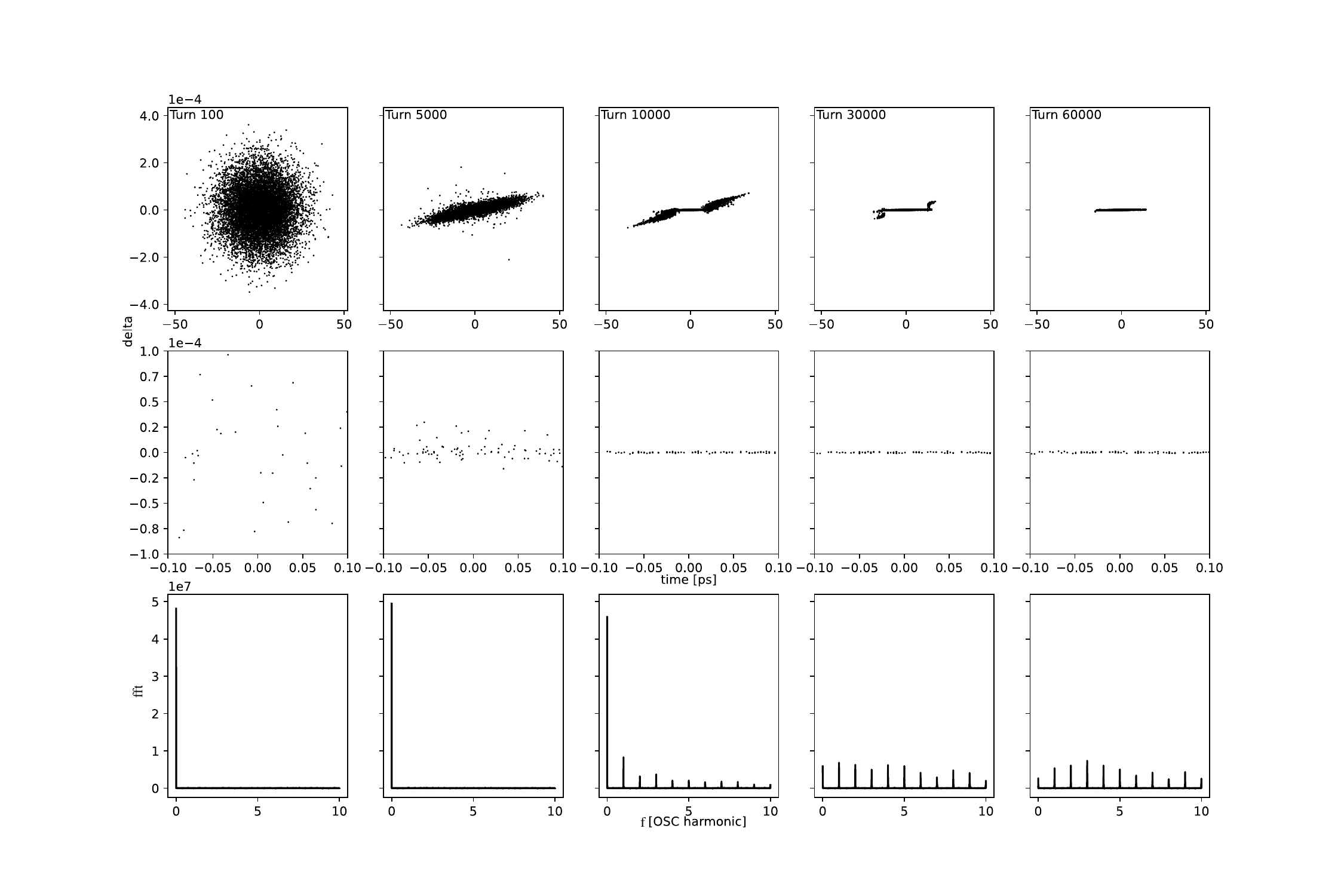}
  \caption{Snapshots from a simple 1D simulation demonstrating the formation of microstructure at the optical wavelength.  The top two panels show the longitudinal phase space (at different scales), and the spectral power density of the longitudinal distributions is shown in the lower panels.}
  \label{fig:OSXDemonstration}
\end{figure}

Figure~\ref{fig:OSXDemonstration} presents a basic 1D simulation of the microbunch formation process in an OSX system.  The OSX mechanism has additionally been validated in preliminary high-fidelity 3D simulations using the \textsc{elegant} simulation package incorporating the OSC model developed alongside the experimental program at IOTA~\cite{Dick2024}, with qualitatively similar results to the longitudinal-only model presented here.  Work is ongoing to include the impact of collective effects within the beam bunch, for example coherent synchrotron radiation and intrabeam scattering, which may be important for an experimental demonstration and eventual applications.

\section{SSMB DEMONSTRATION AT IOTA}

The OSC program at IOTA has been organized to enable a staged development of the technology.  The first phase focused on the experimental demonstration of OSC without optical amplification and was successfully completed in 2021.  The next phase, planned for 2025, will incorporate a high-gain ($30$-$40$\,dB) optical amplifier with turn-by-turn configurability.  The system is under development at the FAST laser lab and has been operated at 30\,dB gain, with further improvements expected~\cite{JarvisIPAC2024}.  Low-alpha operation of the IOTA storage ring has recently been shown experimentally for the first time as part of this program~\cite{WallbankIPAC2024}.  Preliminary simulations suggest that this amplified OSC system within a ring at the demonstrated compactions will be capable of realizing OSX-based SSMB at IOTA.

Conceptual and hardware designs for the OSC system and various operational configurations, including OSX, are under active development.  In parallel, more detailed simulations are ongoing and will include finalized lattice configurations and relevant collective effects.  Additionally, advanced beam control techniques are being developed that integrate reinforcement-learning algorithms into OSC/OSX operations, which will implement the required control policies through a combination of extensive simulations and online training.

\section{CONCLUSION}

Optical Stochastic Crystallization, a novel technique to produce steady-state microbunching within a storage ring bunch based on the mechanism of optical stochastic cooling, has been described.  Initial simulations were performed to study OSX and to aid in the design of an upcoming experimental demonstration at the IOTA ring.  The required low compaction operation of the ring was verified experimentally in 2023, and sufficient performance of the optical amplifier has been demonstrated in a prototype system at the FAST laser lab.

\section{ACKNOWLEDGMENTS}

This manuscript has been authored by Fermi Research Alliance, LLC under Contract No. DE-AC02-07CH11359 with the U.S. Department of Energy, Office of Science, Office of High Energy Physics.

%
%
\ifboolexpr{bool{jacowbiblatex}}%
	{\printbibliography}%
	{%
	
	
} 

%
%


\end{document}